\documentclass[10pt]{article}

\usepackage{amsmath}
\usepackage{amssymb}

\usepackage{graphicx}

\usepackage{cite}

\usepackage{color} 

\usepackage[labelfont=bf,labelsep=period,justification=raggedright]{caption}

\makeatletter
\renewcommand{\@biblabel}[1]{\quad#1.}
\makeatother

\date{}

\begin{document}

\begin{flushleft}
{\Large
\textbf{Irregular dynamics in up and down cortical states}
}
\\
Jorge F. Mejias$^{1 \ast}$,
H. J. Kappen$^2$,
Joaquin J. Torres$^3$
\\
$^1$Centre for Neural Dynamics, University of Ottawa, K1N 6N5 Ottawa, Canada.
\\
$^2$Donders Institute for Brain, Cognition and Behaviour, Radboud University of Nijmegen, 6525 EZ Nijmegen, The Netherlands.
\\
$^3$Department of Electromagnetism and Matter Physics, University of Granada, Fuente Nueva s/n, 18071 Granada, Spain.
\\
$\ast$ E-mail: jmejias@onsager.ugr.es
\end{flushleft}

\section*{Abstract}

Complex coherent dynamics is present in a wide variety of neural systems. A typical example is the voltage transitions between up and down states observed in cortical areas in the brain. In this work, we study this phenomenon via a biologically motivated stochastic model of up and down transitions. The model is constituted by a simple bistable rate model, where the synaptic current is modulated by short-term synaptic processes which introduce stochasticity and temporal correlations. A complete analysis of our model, both with mean-field approaches and numerical simulations, shows the appearance of complex transitions between high (up) and low (down) neural activity states, driven by the synaptic noise, with permanence times in the up state distributed according to a power-law. We show that the experimentally observed large fluctuation in up and down permanence times can be explained as the result of sufficiently noisy dynamical synapses with sufficiently large recovery times. Static synapses cannot account for this behavior, nor can dynamical synapses in the absence of noise.

\section*{Author Summary}

The neural activity observed in most cortical areas often presents a highly complex coherent dynamics. A prominent example is the transitions between two well-defined voltage levels measured in different cortical regions, that is, the up (high activity) and down (low activity) cortical transitions. Although it is well known that the duration of up states is highly irregular (ranging from a few milliseconds to seconds), the origin of such irregularity is still unclear. In this work we propose that the irregularity in the duration of the up states emerges as a consequence of the interplay between synaptic stochasticity and short-term plasticity mechanisms. We show, employing analytical treatments and numerical simulations, that such interplay induces the appearance of power-law distributions of the permanence times in the up state, which may explain the irregularity observed in experiments. On the contrary, such behavior cannot be obtained with static synapses, nor dynamic synapses in absence of noise.

\section*{Introduction}

Neural systems, even in the absence of external stimuli, can exhibit a wide variety of coherent collective behaviors, as {\em in vivo} and {\em in vitro} experiments show~\cite{steriade93c,aertsen96,sanchezvives00}. One of the most prominent examples is the spontaneous transition between two different voltage states, namely up and down states, observed in simultaneous individual single neuron recordings as well as in local field measures. Such behavior, which is generated within the cortex, may provide a framework for neural computations \cite{mccormick05b}, and could also coordinate some sleep rhythms into a coherent rhythmic sequence of recurring cortical and thalamocortical activities \cite{sanchezvives00,steriade93a,steriade93b}. The phenomenon of {\em up and down transitions} has been measured in a number of situations, such as in the primary visual cortex of anesthetized animals \cite{ferster99,ferster00}, during slow-wave sleep \cite{steriade93a,steriade93b,steriade93c}, in the somatosensory cortex of awake animals \cite{petersen03}, or in slice preparation under different experimental protocols \cite{sanchezvives00,Yuste03nature,mccormick03nature}, to name a few. The origin of such structured neuronal activity is still unclear, although several studies have shown that both intrinsic cell properties \cite{tank04,sompolinsky05,abbott07} and the high level of recurrency present in actual neural circuits \cite{sanchezvives00,timofeev01,tsodyks06} may contribute to the generation of up and down transitions. In particular, the contribution that reverberations in recurrent neural networks may have in the appearance of these transitions could depend strongly on synaptic properties. It is known, for instance, that excitatory synapses with a slow dynamics (such as synapses mediated by NMDA receptors) may play a relevant role in the generation of persistent activity or up cortical states \cite{wang99}. On the other hand, several modeling studies indicate that activity-dependent synaptic mechanisms, such as short-term synaptic depression and facilitation, can induce voltage transitions between up and down neural states as well \cite{torresNC,tsodyks06,cortes07,tsodyks08}.

Many crucial points about the understanding of up and down transitions are, however, still lacking. For instance, {\em in vivo} experiments in the cat visual cortex show that the permanence times in the depolarized or up state present a high variability, and can range from a scale of milliseconds to seconds \cite{ferster99}. A similar level of irregularity has also been recently found in {\em in vivo} recordings of up-down transitions in the rat auditory cortex \cite{deco09}, as well as in sleep-wake transitions \cite{timofeev01,pearlmutter09,lo04}, where power-law distributions in the duration of wake states have been measured. Such complexity in the time series of the neuron membrane potentials remains far to be explained, and could reflect scale invariance in permanence times, which could in turn be a (preliminar) indicative of criticality. In fact, there are many recent studies that have shown criticality in different contexts in the brain \cite{chialvo05,beggs03}, as well as in neural network models which present self-organization and criticality properties \cite{pipa07,kappen09,pipa09}, and even it has been reported to occur in sleep-wake transitions in {\em in vivo} conditions \cite{pearlmutter09,lo04}. Although it is worth noting that the irregularity of the dynamics of up and down states is not a sufficient condition for criticality, a concrete characterization of such irregularity may be a convenient starting point for future works on this topic.

To study in detail the relevant issue of irregular up and down cortical dynamics, we propose in this work a minimal model for up and down transitions in neural media. We consider a simple bistable rate model whose stable solutions represent two possible voltage states of the mean membrane potential of the network. More precisely, such states correspond, respectively, to high and low levels of activity in the network (that is, the up and down cortical states). In addition, we consider that the synaptic connections between neurons of the network present short-term synaptic depression (STD) mechanisms, which introduce temporal correlations, as well as synaptic stochasticity, in the dynamics of the system \cite{abbott97,zador97,stevens97,zador98}. A complete analysis of this simple mathematical model depicts, both numerically and within a theoretical probabilistic approach, the appearance of power-law dependences in the distribution of permanence times in the up state. Our results show that the appearance of such scale free distributions is due to the complex interplay between several factors including synaptic stochasticity and the temporal correlations introduced by STD. The emergence of power-law dependences could, indeed, explain the high variability in permanence times in the up state suggested by experiments \cite{ferster99,deco09}.

\section*{Methods}

Our starting point is a bistable rate model, which mimics the dynamics of the electrical activity of a population of interconnected excitatory neurons (although it can be easily extended to other situations) with two stable levels of activity. The model has the form \cite{cowan72}

\begin{equation}
\tau_{\nu} \frac{d\nu (t)}{dt}=-\nu(t) +\nu_m {\cal S} [Jx(t)\nu(t)-\theta]+\zeta(t),
\label{rate}
\end{equation}
where $\nu (t)$ is the mean firing rate of the (homogeneous) neural population, $\nu_m$ is the maximum level of activity which can be reached by the population (in absence of noise), $J(>0)$ is the synaptic coupling strength in absence of STD, and $\theta$ is the firing threshold of the neurons in the population. The variable $\zeta (t)$ is a Gaussian white noise of zero mean and standard deviation $\delta$, which takes into account the inner stochasticity of the neural population (caused by other sources of uncontrolled noise in the system). The parameter $\tau_{\nu}$ is the population time constant, which may be assumed to be around the duration of the synaptic current pulse \cite{gerstner00,gerstnerB}. For generality purposes, we set $\tau_{\nu}=1$, and therefore time and frequency are given in units of $\tau_{\nu}$ and $\tau_{\nu}^{-1}$, respectively. The term ${\cal S}(z)\equiv \frac{1}{2}[1+\tanh(z)]$ represents the transduction function, which gives the nonlinear effect that the mean postsynaptic current (coming from recurrent connections of the neural population) induces in the network mean firing rate. Employing this form for ${\cal S} (z)$, the up and down stable levels of activity correspond to $\nu \simeq \nu_m$ and $\nu \simeq 0$, respectively. 

On the other hand, the variable $x(t)$ in equation (\ref{rate}) takes into account the dynamical modification of the strength of the synaptic connections during short time scales due to high network activity, and it is usually named short-term synaptic plasticity. Based on the model proposed in \cite{abbott97,tsodyks97} for short-term depression, and following previous studies concerning the dynamics of neural populations \cite{tsodyks06}, we assume that $x(t)$ evolves according to
\begin{equation}
\frac{dx (t)}{dt}=\frac{1-x(t)}{\tau_r}-ux(t)\nu(t)+\frac{D}{\tau_r} \xi(t),
\label{std}
\end{equation}
where $\tau_r$ is the characteristic time scale of the STD mechanism, and $u$ is a parameter related with the reliability of the synaptic transmission. According to experimental measurements for these parameters in the somatosensory cortex of the rat \cite{tsodyks97}, we set $\tau_r=1000$ and $u=0.6$ unless specified otherwise\footnote{Assuming a population time constant of $\tau_{\nu}=1~ms$, which would approximately correspond to the duration of a fast synaptic current pulse mediated by AMPA receptors, we obtain $\tau_r=1000~ms$, which is within the physiological range measured in \cite{tsodyks97}.}. The last term on the right hand side of equation (\ref{std}) is added to the original model in \cite{tsodyks97} to include some level of stochasticity in this, otherwise, deterministic description of synaptic transmission. The inclusion of such term constitutes a simple manner of considering the stochasticity due, for instance, to the unreliability of synaptic transmission \cite{stevens97,zador98}, the stochastic properties of receptor-transmitter interactions \cite{korn92}, the sparse connectivity of cortical circuits \cite{kandelB,young95}, or other sources of noise not yet considered (see the Discussion Section for more details). The parameter $D$ controls the strength of this fluctuating term, and $\xi(t)$ is a Gaussian white noise with zero mean and variance one. 



Equations (\ref{rate}) and (\ref{std}) constitute our minimal model of an excitatory neural network with stochastic depressing synapses. The simplifications assumed by such model allows to obtain some analytical derivations for the quantities of interest, and concretely for the probability distributions of permanence times in the up state, denoted by $P(T)$. Bistable systems in the presence of different sources of noise have been theoretically studied in detail in many works \cite{hanggi94,hanggi95,ping06,gaojie08,grinstein05}. Here, however, we have employed a probabilistic approach which is very appropriate for the computation of the distribution of permanence times. In the following, we will derive an approximate expression for $P(T)$ within this approach. First, we obtain the potential function and the conditions in which the dynamics of the system is driven by the variable $x$. After that, we compute the probability distribution of ruin times of $x(t)$ which, as we will see, leads to the probability distribution of permanence times in the up state, namely $P(T)$.

\bigskip
\textit{A. The potential function} 
\bigskip

In order to compute the potential function of the dynamics (\ref{rate},\ref{std}) (namely $\Phi (\nu,x)$) one can see that, for realistic values of $\tau_r$, the dynamics of $x$ is very slow compared to that of $\nu$. We therefore can write equation (\ref{rate}) as
\begin{equation}
\begin{array}{l}
\displaystyle \dot{\nu} =  \displaystyle -\partial_{\nu} \Phi(\nu,x) + \zeta(t)
\\\\
\displaystyle \Phi (\nu,x) =  \displaystyle \frac{1}{2}\nu(\nu-\nu_m)-\frac{\nu_m}{2Jx}\log \cosh (Jx\nu-\theta),
\end{array}
\label{potential}
\end{equation}
where we have adiabatically eliminated $x$ from the dynamics of $\nu$. The extrema of $\Phi$ are given by the solutions of the equation
\begin{equation}
\nu= \frac{1}{2} \nu_m [1+\tanh (Jx \nu-\theta)]\equiv f(\nu)
\label{mfnu}
\end{equation}
In the following, we choose $\theta=J x_0 \nu_0$, with 
$\nu_0 \equiv \frac{1}{2} \nu_m$ and $x_0 \equiv 1/(1+u \tau_r
\nu_0)$. 
With this choice, one can easily check from equation (\ref{potential}) that the potential becomes symmetric in $\nu$ around $\nu_0$ when $x \simeq x_0.$

Equation (\ref{mfnu}) may have one or three solutions, depending on the slope of the hyperbolic tangent and on the value of $\theta$. In order to obtain three solutions of (\ref{mfnu}) (that is, the bistable regime) the maximal slope of the hyperbolic tangent must be large enough, concretely the condition $J x \nu_0>1$ must be fulfilled. In addition, the threshold term must be not too small or too large so that $f(\nu)$ has three crossing points with the straight line $\nu$ rather than one. This last condition can be written, as a first approach, as $f(\nu_1)>\nu_1$ and $f(\nu_2)<\nu_2,$ where $\nu_{1,2}$ are the values where the curvature of the hyperbolic tangent is maximal and minimal, respectively. The points $\nu_{1,2}$ can be easily computed from the third derivative of $f(\nu)$:
\begin{equation}
f'''(\nu)=-\nu_m J^3 x^3 \frac{1-3 \tanh^2 (Jx\nu-Jx_0 \nu_0)}{\cosh^2 (Jx\nu-Jx_0 \nu_0)}.
\end{equation}
By setting $f'''(\nu)=0$ we obtain
\begin{equation}
\nu_{1,2}=\frac{\nu_0 x_0}{x} \pm \frac{\tanh^{-1} (\sqrt{1/3})}{Jx}.
\end{equation}
Using now these values for $\nu_{1,2}$, the conditions $f(\nu_1)>\nu_1$ and $f(\nu_2)<\nu_2$ can be written as

\begin{equation}
-\nu_0 \sqrt{1/3} +\frac{1}{Jx} \tanh^{-1} (\sqrt{1/3})< \frac{\nu_0
  x_0}{x}-\nu_0 < \nu_0 \sqrt{1/3} -\frac{1}{Jx} \tanh^{-1} (\sqrt{1/3}),
\label{eq7}
\end{equation}
which implies that, in order to have one maxima and two minima in $\Phi (\nu,x)$, the variable $x$ must be in the range $x_1 < x < x_2$, where

\begin{equation}
x_1 \equiv \frac{\nu_0 x_0+\frac{1}{J} \tanh^{-1}(\sqrt{1/3})}{\nu_0 (1+\sqrt{1/3})},
~~~~~~~~~~
x_2 \equiv \frac{\nu_0 x_0-\frac{1}{J} \tanh^{-1}(\sqrt{1/3})}{\nu_0 (1-\sqrt{1/3})}.
\label{xrange}
\end{equation}
From equation (\ref{xrange}), one can see that the range of $x$ that allows to have three extrema in the potential is
\begin{equation}
\Delta x \equiv x_2-x_1= \sqrt{3} ~x_0 -\frac{3}{J \nu_0} \tanh^{-1} (\sqrt{1/3}).
\end{equation}
The condition $\Delta x>0$ implies $J x_0 \nu_0 \gtrsim 1.14$ which is, therefore, a sufficient condition to obtain a double well potential\footnote{One can find, however, a small discrepancy between our approximate prediction and the actual properties of $\Phi (\nu,x)$. The discrepancy appears because we have assumed that a sufficient condition for the existence of the three fixed point solutions of equation (\ref{mfnu}) is that $f(\nu_1)>\nu_1$ and $f(\nu_2)<\nu_2$, and such assumption is only approximately correct. Plotting directly the potential as a function of $\nu$ reveals that the condition to obtain a double well potential for $x \simeq x_0$ is $J x_0 \nu_0 >1,$ rather than $J x_0 \nu_0 >1.14$.} for some value of $x$. Assuming that this condition is satisfied, three different shapes for the potential function $\Phi (\nu,x)$ can be found, as the figure 1A illustrates. When $x<x_1$ the potential function presents only one minimum, located around $\nu \simeq 0$. Similarly, for $x_2<x$ the potential presents also a single minimum, but now located around $\nu \simeq \nu_m$. Finally, for $x_1<x<x_2$ the potential will take a double well shape, with the maximum being located around $\nu \simeq \nu_0$ and the minima located around $\nu \simeq 0$ and $\nu \simeq \nu_m$, respectively. 

It is worth noting that $x_1<x_0<x_2$, with $x_0$ being the mean value of $x$. Due to this, if the range $\Delta x$ is small compared with the fluctuations of $x$, namely $\sigma_x$, the potential function will spend most of the time in the regimes $x<x_1$ and $x_2<x$, with the double well regime appearing only when the system tries to jump from one of these regimes to the other (that is, when $x \simeq x_0$). A direct consequence of this is that the mean firing rate will be basically switching between the up and down states (that is, $\nu \simeq 0$ and $\nu \simeq \nu_m$), and that this switching will be driven by the dynamics of $x$, as the figure 2 illustrates. Therefore, one expects that the distribution of permanence times of $\nu$ in the up (down) state, becomes approximately equal to the distribution of permanence times of $x$ in the $x>x_0$ ($x<x_0$) regime, as long as $\Delta x \ll \sigma_x$ is satisfied\footnote{It should be noted that, since $x$ is a fraction of available neurotransmitters, its value should be kept within the range $[0,1]$. In practice, this means that the value of $\sigma_x$ must not be too large, so in order to make $\Delta x \ll \sigma_x$ one has to restrict to $\Delta x$ small. In the results presented here, $x$ remain in its realistic range of values, and imposing {\em ad hoc} restrictions in such a way that $x$ is always within the range $[0,1]$ does not affect the results obtained here.}. Due to this equivalence, in order to compute $P(T)$ we only need to compute the distribution of permanence times of the variable $x$ in the $x>x_0$ regime, denoted as $P_{x}(T)$.

\bigskip
\textit{B. Distribution of permanence times} 
\bigskip

In order to compute the distribution of permanence times of $x$ in the $x>x_0$ (or $x<x_0$) regime, one can assume that the firing rate takes its mean value $\nu \simeq \nu_0$ in equation (\ref{std}). This is a reasonable approach since $x$ is much slower than $\nu$ for realistic values of the parameters. Considering this approach, and after the rescaling $z \equiv (1+u \tau_r \nu_0)x-1$, equation (\ref{std}) can be written as
\begin{equation}
\frac{dz(t)}{dt}=-\frac{z(t)}{\tau}+\frac{D}{\tau} \xi(t)
\label{OU}
\end{equation}
which is the equation of the Ornstein-Uhlenbeck (OU) process (see \cite{vankampenB} for details), with $\tau \equiv \tau_r/(1+u \tau_r \nu_0)$ being the correlation time and $z_0 \equiv z(x_0)=0$. Therefore, computing the distribution of permanence times in the up state for our system is equivalent to obtain the distribution of the so called {\em ruin times}\footnote{If we consider a stochastic process $y(t)$ starting at $t=t_0$ from $y=y_0$, the ruin time is defined as the interval $t_1-t_0$, where $t_1$ is the time at which $y(t)$ returns to $y_0$ for the first time. Since $y(t)$ is a stochastic process, the ruin times are stochastic quantities which follow a certain probability distribution.} for the OU process \cite{chandrasekhar43,newman05}. The strategy employed here to calculate the distribution of ruin times is based on the relation between the ruin time and the {\em first passage time}, which is the typical time that a stochastic process needs to arrive at a certain threshold value when starting from a certain initial condition \cite{newman05}. Because of the symmetry of the OU process, the distribution of ruin times are equivalent when considering excursions of the variable $z$ in the $z<0$ region or in the $z>0$ region. If we consider excursions in the $z<0$ region, we can set a small positive threshold $\epsilon$ near zero (that is, $0< \epsilon \ll 1$), in such a way that the typical ruin time will be approximately equal to the corresponding first passage time, as the figure 1B illustrates. The excursions in the region $z>0$ typically lead to very short first passage times (since $\epsilon$ is too small) which we will not take into account in our calculations by considering only large enough ruin times.


The first passage time for the OU process with a small threshold $\epsilon$ can be performed by using the relation

\begin{equation}
{\cal P}(\epsilon,T|0,0)=\int_0^T dt ~{\cal P}(\epsilon,T|\epsilon,t) ~\rho (t),
\label{fpt}
\end{equation}
where ${\cal P}(a,t_a|b,t_b)$ is the conditional probability distribution of the OU process, and $\rho(t)$ is the first passage time distribution. This equation can be solved by taking into account the following property of the Laplace transformation
\begin{equation}
f_1(t)=\int_0^t dt'~f_2(t-t')~f_3(t') ~~ \Longrightarrow ~~ \hat{f_1}(s)=\hat{f_2}(s)~\hat{f_3}(s),
\label{laplace}
\end{equation}
where $\hat{f_i}(s)$ is the Laplace transform of $f_i(t)$. By solving the Fokker-Planck equation associated with equation (\ref{OU}), one can obtain the conditional probability for the OU process
\begin{equation}
{\cal P}(z_2,t_2|z_1,t_1)=\frac{1}{\sqrt{2 \pi \sigma_x^2 [1-\exp(-2 \Delta
      t/\tau)]}} \exp \left \{ -\frac{[z_2-z_1 \exp(-\Delta t/\tau)]^2}{2
  \sigma_x^2 [1-\exp(-2 \Delta t/\tau)]} \right \}
\label{OUcondprob}
\end{equation}
where $\Delta t \equiv t_2-t_1>0$, and $\sigma_x \equiv D/\sqrt{2 \tau}$ being the standard deviation of $x.$ From expression (\ref{OUcondprob}), and assuming that $\tau$ is large enough\footnote{More precisely, we assume that $\tau \gg \Delta t$, which is a valid hypothesis since most of the permanence times in the up state are much lower than $\tau$.}, one arrives at

\begin{equation}
\begin{array}{ll}
{\cal P}(\epsilon,T|0,0) \simeq \frac{1}{\sqrt{4 \pi \sigma_x^2 ~T/\tau}} \exp \left( -\frac{\epsilon^2 \tau}{4 \sigma_x^2 T}\right) 
\\\\
{\cal P}(\epsilon,T|\epsilon,t') \simeq \frac{1}{\sqrt{4 \pi \sigma_x^2 ~(T-t')/\tau}} \exp \left( -\frac{\epsilon^2 (T-t')}{4 \sigma_x^2 \tau}\right).
\end{array}
\end{equation}
We denote $f_1(T) \equiv {\cal P}(\epsilon,T|0,0)$ and $f_2(T-t') \equiv {\cal P}(\epsilon,T|\epsilon,t')$. Employing the Laplace transformation in $f_1 (T)$ and $f_2 (T-t')$ the following expressions are obtained

\begin{equation}
\begin{array}{ll}
\hat{f_1}(s)=\sqrt{\frac{\tau}{4s \sigma_x^2}} \exp \left( -\sqrt{\epsilon^2 \tau s/ \sigma_x^2}\right) 
\\\\
\hat{f_2}(s)=\tau/\sqrt{\epsilon^2+4s\tau \sigma_x^2}.
\end{array}
\end{equation}
Now, taking into account the property (\ref{laplace}) in equation (\ref{fpt}), the expression for $\hat{\rho} (s)$ is
\begin{equation}
\hat{\rho} (s)= \sqrt{\frac{\epsilon^2+4s\tau \sigma_x^2}{4s \tau \sigma_x^2}} \exp \left( -\sqrt{\epsilon^2 \tau s/\sigma_x^2}\right).
\label{casicasi}
\end{equation}
Finally, for small $\epsilon$ one can approximate $\epsilon^2+4s\tau \sigma_x^2 \simeq 4s\tau \sigma_x^2$. With this approximation, one can easily perform the inverse Laplace transformation to equation (\ref{casicasi}) and obtain the distribution of first passage times for the OU process

\begin{equation}
\rho (T)=\sqrt{\frac{\epsilon^2 \tau}{4 \pi \sigma_x^2}}~ T^{-3/2} \exp \left( -\frac{\epsilon^2 \tau}{4 \sigma_x^2 T}\right).
\end{equation}
A similar expression may be obtained if one considers more classical derivations of the first passage time of the OU process (see, for instance, \cite{alili05}). In order to obtain the distribution of ruin times of the OU process, one has to consider a small (but positive) value of $\epsilon$, which leads to $\rho (T) \sim T^{-3/2}$. The distribution of ruin times of the variable $x$, namely $P_{x}(T)$, and therefore, the distribution of permanence times in the up state, namely $P(T),$ for our system are also given by
\begin{equation}
P(T)\sim T^{-3/2},
\label{ptd}
\end{equation}
which corresponds to a power-law probability distribution for $T$. 

Summarizing, the three following conditions must be fulfilled to obtain a power-law dependence in $P(T)$ with exponent $-3/2$:

\begin{itemize}
\item Large enough values of $\tau_r$. With this condition, we ensure that the dynamics of $x(t)$ is much slower than that of $\nu(t)$.

\item Large enough values of $D$. In particular, we must have $D\gg2 \tau \Delta x$, according to the condition $\Delta x \ll \sigma_x$ and the definitions $\sigma_x \equiv D/\sqrt{2 \tau}$ and $\tau \equiv \tau_r/(1+u \tau_r \nu_0)$. This condition can be achieved even with very small values of $D$, since $\Delta x$ can be arbitrarily small (by increasing $J$, for instance).

\item The condition $J x_0 \nu_0 >1$ must hold to ensure the existence of two well defined (up-down) states.
\end{itemize}

All these conditions may be easily achieved (up to some point) with realistic values of the model parameters, indicating that power-law distributions of the permanence times in the up state are plausible to be found in actual cortical media.


\section*{Results}

As we have stated in the previous sections, equations (\ref{rate}-\ref{std}) govern the dynamics of our simplified neural system. A typical time series of the dynamics of this model, for the case of deterministic synapses (that is, $D=0$), is depicted in figure 2A. In this case, the mean firing rate of the population is characterized by a periodic switching between up and down states. This type of periodic behavior was already found and analyzed in previous theoretical studies \cite{torresNC,tsodyks06,abbott07} and yields bimodal histograms for the mean firing rate of the neural population (see figure 2B), as the experiments indicate \cite{sanchezvives00}. However, these approaches ignore the stochastic nature of synaptic transmission, and other forms of stochasticity at the synaptic level, which seem to be crucial for information processing in neural systems~\cite{stevens97,zador98,delarocha05}. Considering a certain level of synaptic stochasticity in addition to STD in our model, one obtains a qualitatively different emergent behavior, as is shown in figure 2C for $D=20$. The mean firing rate presents then a complex switching between up and down states, and in particular involves a high variability in the permanence times in the up state.

When deterministic synapses are considered (that is, $D=0$) the dynamics of the mean firing rate becomes quasi periodic, as it was reported in \cite{torresNC,tsodyks06, cortes07}, for instance. This type of dynamics naturally leads to exponential distributions for the permanence times\footnote{More precisely, for $D=0$ our model is similar (except for the term $\zeta(t)$) to the one analyzed in \cite{tsodyks06}, which shows periodic oscillations of the network mean firing rate. In the case of our model with $D=0$, the term $\zeta (t)$ introduces certain level of stochasticity which turns these periodic oscillations into quasi-periodic oscillations. This leads to the exponential distributions for the permanence times in the up state.}. When $D$ is increased, however, the stochasticity of the synapses leads to the appearance of power-law distributions for the permanence times in the up state. This behavior is shown in figure 3A, where low values of $D$ corresponds to exponential distributions for $P(T)$, while larger values of $D$ give $P(T) \sim T^{-3/2}$ as predicted by our theoretical calculations. Such power-law distributions may explain the high variability of permanence times in the up state, which has been observed in a number of {\em in vivo} experiments, such as in the cat visual cortex \cite{ferster99} and rat auditory cortex \cite{deco09}, to name a few. Interestingly, similar power-law dependences have been observed during sleep-wake transitions {\em in vivo} when one measures the distribution of permanence times in the wake state \cite{pearlmutter09,lo04}. On the other hand, exponential-like distributions, obtained for the case of having $D=0$, are not able to explain this variability of the duration of up states. 

By looking at the data for $D=20$ in figure 3A, one can observe the existence of a small deviation of the numerical results (blue points) with respect to the theoretically predicted slope (solid line) for very large values of $T$. Such deviation is due to the fact that the separation of timescales between the dynamics of $\nu(t)$ and $x(t)$ (a neccessary condition to obtain power-law dependences in  $P(T)$) is only approximate when considering realistic values of the parameters (and in particular, realistic values for $\tau_r$). More precisely, the approximation fails when the activity of the system falls in the occasional periods of very long permanence times in the up state (that is, for large enough values of $T$, comparable with $\tau_r$). In order to study the effect of the separation of timescales between $\nu(t)$ and $x(t)$, we have computed $P(T)$ for different (increasing) values of $\tau_r$ while keeping fixed values for $\Delta x$ and $D/\tau_r$ (this can be done by properly modifying $J$ and $D$ with $\tau_r$, respectively). As a consequence of this, the only effect of increasing $\tau_r$ will be a clearer separation of timescales between $\nu(t)$ and $x(t)$. The results are shown in figure 3B, where one can see that larger values of $\tau_r$ (that is, clearer separation of timescales) lead to a displacement of the effective cut-off towards higher values of $T$, as expected, and a clearer power-law distribution emerges. 

It is worth noting that the appearance of an effective cut-off in $T$ for realistic conditions does not represent an unrealistic feature of the model, but rather it constitutes a prediction about the effective range of permanence times which are expected to occur in actual neural systems. Indeed, for realistic values of the parameters, our results predict permanence times in the up state up to $\sim 1000~ms$, which is the maximum permanence time observed in experimental realizations \cite{ferster99}. Larger permanence times in the up state (of about $10$ seconds, for instance) should be expected to appear only as a consequence of input driven mechanisms (such as persistent activity associated with working memory tasks \cite{fuster71,rakic95}), and not as a consequence of spontaneous transitions between different voltage levels, which are the matter of interest in this work. 

For a better characterization of the dynamics of the system, one can use, for instance, other statistical magnitudes such as the autocorrelation function $C(t')$ of $\nu,$ which can be defined as
\begin{equation}
C(t') \equiv \left\langle \nu(t+t') \nu(t)-\nu(t)\nu(t') \right\rangle .
\label{autocorr}
\end{equation}
Here, $\left\langle \cdots \right\rangle $ indicates a temporal average. The autocorrelation function is depicted in figure 4A for the case of deterministic depressing synapses ($D=0$) and stochastic depressing synapses ($D=20$). $C(t')$ presents, for $D=0,$ two well located peaks at $t'\simeq \pm 200$, which indicates a strong periodicity of the time series (as can be seen in figure 2A). On the contrary, the inclusion of a certain level of intrinsic stochasticity in the dynamics of $x$ introduces more pronounced temporal correlations in the dynamics of the system. This fact reflects the existence of long permanence stays in the up state, which occurs with more probability for high enough values of $D,$ as we have already discussed. 


The spectral properties of the dynamics can be analyzed as well, via the power spectrum defined as 

\begin{equation}
F(f) \equiv \int C(t') \exp (2 \pi i f t' ) dt' .
\label{powerspectra}
\end{equation}

As one could expect, the power spectrum of the case $D=0$ presents a pronounced peak around a certain frequency, which in the particular case presented in the figure 4B is $f \sim 5\cdot 10^{-3}$. The power spectrum for higher values of $D$ shows however different properties than the case $D=0$. For instance, the figure 4B (which considers $D=20$) indicates an approximated power-law behavior for the power spectrum, $F(f) \sim f^{-\beta}$ with $\beta \simeq 1.7$. This scale-free dependence can be understood by considering that, if $P(T)$ is algebraic with exponent $\gamma$, the corresponding power spectrum becomes also algebraic with exponent $\beta$, where the equation $\gamma+\beta=3$ relates both exponents \cite{grinstein05}. In our particular case, since $\gamma \simeq 1.5$, one obtains a theoretical prediction of $\beta \simeq 1.5$ for the exponent of the power spectrum. The theoretical relation between $P(T)$ and $F(f)$ exposed above, however, is only valid under the so called {\em single interval approximation}, which implies that the integration variable $t$ in equation (\ref{powerspectra}) is smaller than the permanence time $T$ (see \cite{grinstein05} for details). This condition does not strictly hold for our system (where $T$ ranges over several scales), and therefore it may introduce deviations in the theoretically predicted value of $\beta$ (which is around $\beta \simeq 1.5$) with respect to the value found in simulations (of around $\beta \simeq 1.7$). 


Besides the level of synaptic stochasticity, i.e. $D$, other parameters of the model could have an important effect on the dynamics as well. The parameter $\delta$, for instance, controls the level of stochasticity of the dynamics of $\nu$, and therefore one should expect that increasing its value could strongly influence the probability distribution $P(T)$. This is shown in figure 5A, where an increase of $\delta$ disrupts the appearance of power-law dependences, and exponential distributions appear instead. This change in $P(T)$ is due to the fact that high levels of the additive noise $\delta$ make the system to jump more frequently from one state to the other, and therefore long stays in the up state (and thus distributions with long power-law tails) rarely occur. 

The parameters involving the dynamics of $x$ also affect the probability distributions $P(T)$. The parameter $u$, for instance, is responsible for the modulation of $x$ via the mean firing rate $\nu$ (see equation (\ref{std})), and therefore it can influence both the dynamics of $x$ and $\nu$. As one may see in figure 5B, when $u$ takes low values a bump in $P(T)$ emerges for high $T$. Such deviation from the power-law dependence indicates that long stays in the up state occur more frequently than in the power-law case. Attending at equation (\ref{std}), one can see that an increase of the mean firing rate $\nu$ decreases the variable $x$ via the parameter $u$. Therefore, if $u$ takes lower values the decrement of $x$ will be smaller. As a consequence, the stays of $x$ in the $x_0 \ll x$ regime (see Methods Section) will last longer, and the stays of the system in the up state will also last longer, causing the observed deviation from the power-law tendency. It should be noted, however, that the values of $u$ which allow the appearance of power-law dependences in $P(T)$ for our model agree with the values of $u$ measured in actual cortical media where up and down transitions are observed \cite{tsodyks97}. 

We have also analyzed in detail the effect that varying $\tau_r$ has on the probability distribution of permanence times. Note that, contrarily to the previous study presented above, we have now varied the parameter $\tau_r$ while all the other parameters are kept fixed. This implies that the modification of $\tau_r$ will now have an effect on the separation of timescales between $\nu(t)$ and $x(t)$, but also on the concrete value of $\Delta x$ and on the amplitude of the noisy term of equation (\ref{std}) (namely $D/\tau_r$). The results are shown in figure 5C, where one can distinguish three different regimes as a function of the particular value of $\tau_r$. For low $\tau_r$ (red region in the figure), the probability distributions show an exponential decay for large permanence times. The reason for this decay is that, for low $\tau_r$, the variable $x$ does not perform long excursions in the region $x_0 \ll x$ (see Methods Section), and therefore the probability to have large values of $T$ decreases and the power law behavior for $P(T)$ is not obtained. As $\tau_r$ is increased, long excursions for $x$ begin to occur, and we obtain a power law behavior $P(T)\sim T^{-3/2}$ (green region in the figure). Finally, one can appreciate that, for even larger values of $\tau_r$ (blue region in the figure), the probability distribution of permanence times in the up state presents a power law dependence $P(T) \sim T^{-\gamma(\tau_r)}$ with $\gamma(\tau_r)>3/2,$ being an increasing function of $\tau_r$. Such dependence can not be explained by our previous theoretical predictions, based in the assumption that the system is in the bistable regime, and deserves a detailed analysis which will be exposed in the next section.

\subsection*{Further analysis}


In the Methods Section, we established several conditions which had to be fulfilled in order to obtain power law dependences for $P(T)$. In particular, our previous analysis indicates that the condition $J x_0 \nu_0 >1$ must hold in order to have a potential function $\Phi (\nu,x)$ with three extrema (bistable regime). However, as we will see in the following, power law expressions for $P(T)$ may appear even if the potential function has only one extremum in $\nu$ (concretely, one minimum), although the origin of such power law distributions is different from the one considered in previous sections, as we will see. 

When $J x_0 \nu_0 <1$ (which occurs for $J\ll 1$ or $\tau_r\gg 1$, for instance), the potential function $\Phi (\nu,x)$ has only one minimum in $\nu$, whose location strongly depends on $x$. An approximated expression for the location of this minimum as a function of $x$ can be obtained by expanding the hyperbolic tangent of the fixed point expression of $\nu(t)$ (see equation (\ref{mfnu})) around its argument (which is small in this limit), yielding
\begin{equation}
\nu_{min} \simeq \nu_0 (1-J x_0 \nu_0)+(J \nu_0^2-J^2 x_0 \nu_0^3)x+J^2 \nu_0^3 x^2,
\label{approxnumono}
\end{equation}
where $\nu_{min}$ is the value of $\nu$ which corresponds to the minimum of the potential function. Therefore as $x$ varies around $x_0,$ the location of the minimum of the potential $\nu_{min}$ also varies in the same way around $\nu_0$. As an example, time series of both $\nu$ and $x$ are shown in figure 6A for a given set of parameters which satisfies $J x_0 \nu_0 <1$. In this time series, the variable $\nu$ fluctuates around the value $\nu_{min}$, which is fully determined by $x$ (that is, the variable $\nu$ becomes a slave variable of $x$). The predictions of equation (\ref{approxnumono}) agree approximately well with simulations and with the numerical evaluation of the fixed points of equation (\ref{rate}), as the figure 6B shows.

Since $\nu$ behaves now as a stochastic variable which does not present a clear bistable dynamics, the numerical computation of the distribution of the permanence times will depend on the exact value of $\nu$ above which the system is considered to be in the up state. As we have seen before, this {\em threshold} value takes the form $\eta \nu_m$ (see caption of figure 3), where usually $\eta$ may take a value between $0.6$ and $0.9$. While the results presented for $J x_0 \nu_0 >1$ (that is, the bistable regime) are quite robust for different values of $\eta$, in the regime $J x_0 \nu_0$ this parameter has indeed some effect on $P(T)$, which indicates the difficulty to accurately analyze the up and down dynamics in this case.

In figure 7A, one observes that the distribution $P(T)$ shows also a power law behavior $P(T) \sim T^{-\gamma}$ for $\eta=0.75$ and different values of $D$, for a set of parameter values which satisfies $J x_0 \nu_0<1$ (that is the monostable regime). The concrete value of $\gamma$ depends strongly on $D$ and it has also a weaker dependence with $\eta$, as the figure 7B illustrates. This type of power-law behavior appearing in the monostable regime corresponds to the blue region in figure 5C, as well.


It is worth noting that actual recordings of up and down transitions does not present a clear distinction between up and down states, and several nontrivial methods are commonly employed to discriminate between both states \cite{sanchezvives07}. Therefore, the results found for the regime $J x_0 \nu_0 <1$ could indeed reflect the behavior of actual cortical up-down transitions, showing power law dependences in $P(T)$ with $\gamma >3/2$ and indicating that the concrete nature of the transitions is a synaptic-driven monostable dynamics.


For a complete characterization of the model, one can summarize all the observed behaviors in a phase plot such as the one presented in figure 8A. A total of four different behaviors can be found in the $(\tau_r,~D)$ space. The first one concerns the dynamics of $\nu$ whose permanence times in the up state follows an exponential distribution (labeled as ``E'' in the figure). If the noise amplitude $D$ is sufficiently high, one can increase the value of $\tau_r$ to reach the regime ``C'', in which the dependence $P(T) \sim T^{-1.5}$ is obtained. By increasing $\tau_r$ even more, the probability distribution $P(T)$ takes the form $\sim T^{-\gamma}$, with $\gamma >1.5$ (regime denoted by ``S''), as we have already seen in figure 6. Finally, we also observe that when the depression time scale is not large enough (and $D\lesssim 3$), a regime of quasi-periodic time series of $\nu$ is obtained, with a well-defined duration of up states (regime denoted by ``P''). The lines between the different regimes have been obtained by visual inspection of $P(T)$ for different values of $\tau_r$ and $D$. In particular, the regime ``P'' is characterized by the appearance of a {\em bump} in the probability distribution for some value of $T$ (which reflects a preferred duration of the up state), and the existence of such bump has been used as a criterion to distinguish between regimes ``P'' and ``E''. Similarly, we assumed that the regimes ``C'' and ``S'' correspond to the situation in which a power-law behaviour that extends for two decades or more is found for $P(T)$. Such criterion, together with an estimation of the slope of the power-law via standard Levenberg-Marquardt fitting algorithms, allows to distinguish between regimes ``E'', ``C'' and ``S''. 

It must be clarified, however, that actual up and down cortical transitions present most likely a richer repertoire of dynamical regimes than the one obtained with our simplified model. It is known, for instance, that attractor neural networks with dynamic synapses may exhibit different dynamics corresponding to memory, non-memory and switching regimes \cite{torresNC,cortes07}. In this work, we have extensively explored different regimes of switching behavior, and its implications for the up and down dynamics observed in the cortex. The memory and non-memory regimes, however, can be also found in our simplified model by assuming that $D,~\delta \rightarrow 0$. After taking these limits, the system will be in the memory regime if the potential function $\Phi(\nu,x)$ is bistable, or in the non-memory regime if $\Phi(\nu,x)$ is monostable.

\section*{Discussion}

We have shown that the experimentally observed large fluctuations in up and down permanence times can be explained as the result of sufficiently noisy dynamical synapses with sufficiently large recovery times. Our study suggests that a power-law distribution for these permanence times may emerge as a consequence of these two ingredients. Static synapses cannot account for this behavior, nor can dynamical synapses in the absence of noise. 

The origin of up and down cortical transitions is still unclear, although different factors that may influence their occurrence have been recently reported. It is known, for instance, that inhibitory GABAergic currents strongly contribute to the temporal coding and spike timing precision of cortical networks during up states of activity \cite{sanchezvives00,mccormick05,compte06b}. Several modeling studies also show the relevance of inhibitory interneurons in the generation of many types of oscillations in the brain (see for instance \cite{brunel00}). However, other studies indicate that most of the main features of up and down transitions depends strongly on synaptic plasticity mechanisms, both of long-term and short-term ones \cite{fukai08,tsodyks06}, and that the transitions appear even in the absence of inhibition \cite{tsodyks06}. In this work we have made the common assumption that the effects of inhibition can be treated as additive and can be incorporated in the threshold of the neuron. This is known to be a valid approximation in mean field neural network analysis, but may fail when precise timing and details of the dynamical aspects of the neuron affect the inhibition \cite{rotstein05,claassen08}. 

Regarding to synaptic characteristics, recent works show that synaptic fluctuations could have an important role in the generation of transitions between up and down states \cite{cortes06,abbott07,johnson08}. Since our model introduces stochasticity in the synaptic dynamics in a highly simplified manner, however, the last term in equation (\ref{std}) should not be associated only with ureliability in synaptic transmission. Indeed, we have assumed that other sources of stochasticity may be contributing to this fluctuating term in the mean-field quantities $\nu(t)$ and $x(t)$. For instance, it is widely known that connectivity in actual cortical media is highly sparse. Such feature implies that, in order to obtain the mean-field quantity $x(t)$, the average over synapses must be performed over a number of $C$ synapses, with $C$ ranging over $100 \sim 1000$ connections per neuron \cite{kandelB}. In this situation, the fluctuations of $x(t)$ would be of order $1/\sqrt{C}$, which leads to a range of $0.1 \sim 0.03$ for the values given above. As we have seen, our results state that a value of $D/\tau_r=0.02$ is enough to obtain power-law distributions (see figure 3), which lies within this range. Therefore, topology-induced fluctuations constitute an important source of stochasticity which could be responsible of the appearance of power-law distributions in $P(T)$. Other sources of stochasticity at synaptic level, such as the stochastic properties of receptor-transmitter interactions, may also contribute to the last term of equation (\ref{std}). Moreover, the low activity rates typical from cortical media lead to a poor time-averaring of the incoming input, and therefore the fluctuations at the postsynaptic level will be large at these short-time scales (of the order of the typical synaptic integration time constant). 

On the other hand, the amplitude of the noisy term, $D/\tau_r$, does not need to be very high to induce the appearance of power-law distributions in $P(T)$. As we have stated above, a sparse connectivity already induces a level of stochasticity which is within the desired range, for instance. Furthermore, the noisy term could even be arbitrarily small: attending to our theoretical predictions, a neccessary condition to have power-law distributions is that fluctuations of $x(t)$ must be much larger than $\Delta x$ (see Methods Section). Since $\Delta x$ may be lowered to arbitrary levels (by increasing $J$, for instance), even a small noisy term in the dynamics of $x(t)$ may induce power-law distributions. 

It is also known that short-term synaptic mechanisms, such as short-term depression and facilitation, usually play a role in the efficient processing of information. In particular, they may be relevant in many tasks, such as in signal detection and coding \cite{abbott97,compte06,mejias08jcns,mejias10jcns} or switching between different activity patterns previously stored \cite{cortes07,mejias09neco}. However, their role on the transitions between cortical states has been pointed out only by a few studies \cite{timofeev01,timofeev00,tsodyks06}, and their possible effects on the statistics of the transitions, which is the focus of our work, have been ignored. To the best of our knowledge, the present study is the first one which analyzes, even in a simplified manner, the strong effect of synaptic stochasticity-- in a general sense-- and dynamic synapses in the statistics of the up and down transitions. The possible role of other short-term synaptic mechanisms, such as STF, has not been addressed yet and constitutes a interesting issue still open. 

In our analysis we assumed that the dynamics is symmetric in the up and down states. This is in contradiction with experimental evidences \cite{sebas-mavi} which shows that power-law distributions are obtained for permanence times in the up state, while permanence times in the down state are exponentially distributed. However, this discrepancy disappears when one considers a more realistic transduction function which gives an asymmetric potential for the dynamics, and as a consequence the up-down symmetry is broken. More detailed studies considering, for instance, some of the biologically realistic aspects discussed above, should be performed to test our predictions. In particular, a more elaborated study considering realistic neuron models (such as Hodgkin-Huxley model \cite{hodgkin52}) and stochastic STD models (see \cite{abbott97,delarocha05}, for instance) is necessary, as 
well as more detailed experimental studies which may confirm our predictions.

From a general point of view, evidences of criticality have been recently found in an increasing number of neural systems, such as in the functional connectivity of the living human brain \cite{chialvo05}, in critical avalanches of neuronal activity \cite{beggs03}, or in sleep-wake transitions \cite{lo04}, to name a few. According to the results presented in this work, transitions between up and down cortical states could also present some relevant properties typical of systems at criticality. Some of these properties have been already measured in experiments, such as a high sensitivity of the system to external stimuli \cite{ferster00}, or the presence of power-law dependences in the power spectra of the neural dynamics \cite{mccormick05}. 

It is worth noting that other kind of probability distributions for $P(T)$, such as a log-normal distribution, could also satisfactorily explain the irregularity in the up states found in experiments. Our study shows the importance of some biophysical factors, such as the neurotransmitter recovery time and the inherent synaptic stochasticity, and predicts a power-law dependence on $P(T)$ as a consequence of such factors. However, further study is needed to investigate other mechanisms, not taken in account in this work, which could influence the permanence times in the up state. In a more general sense, our results may proportionate a new perspective of the phenomena of up and down transitions (and a theoretical framework) that could serve to conciliate the main experimental findings, and that could help for a deep understanding of this complex dynamics of the brain activity.

\section*{Acknowledgments}
This work was supported by the \textit{MEyC--FEDER} project FIS2009-08451 and the \textit{Junta de Andaluc\'{\i}a} project P07--FQM--02725. We thank useful discussions with Sebastiano de Franciscis and Samuel Johnson.



\section*{Figures}

\begin{figure}[!ht]
\begin{center}
\includegraphics[width=6.5in]{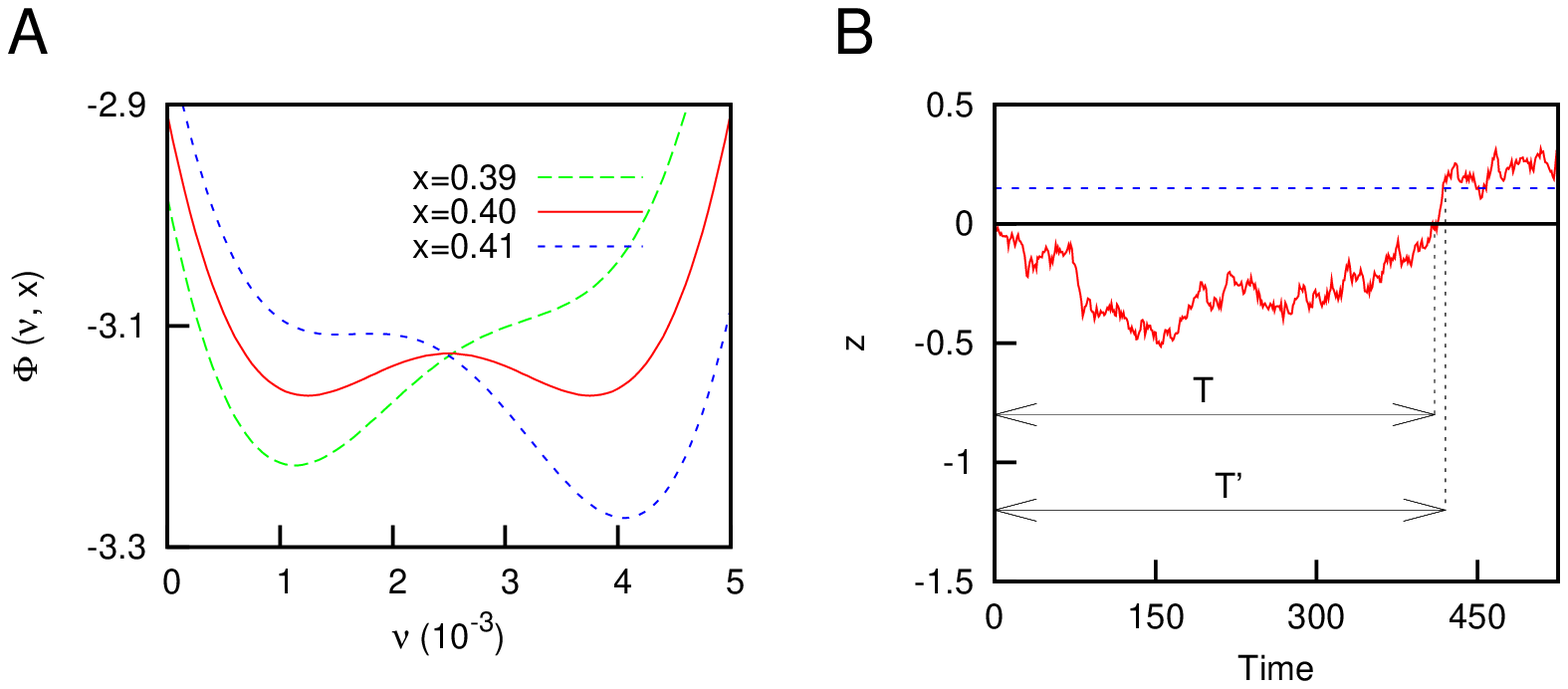}
\end{center}
\caption{
{\bf Considerations for the mean-field approach.} (A) Potential function $\Phi (\nu,x)$, as a function of the mean firing rate $\nu$ and for different values of $x$. One can appreciate the different regimes explained in the main text. Other parameters are $J=1.1~V, ~\tau_r=1000, ~u=0.6$ and $\nu_m=5\cdot10^{-3}$. (B) An Ornstein-Uhlenbeck (OU) process (see equation \ref{OU}) with $\tau=1000$ and $D=20$. A typical return event (with return time $T$) and a first passage event (with first passage time $T'$) are indicated for illustrative purposes. For the first passage time, the threshold (depicted as a blue dashed line) was fixed to 0.15.}
\label{fig1}
\end{figure}

\begin{figure}[!ht]
\begin{center}
\includegraphics[width=6in]{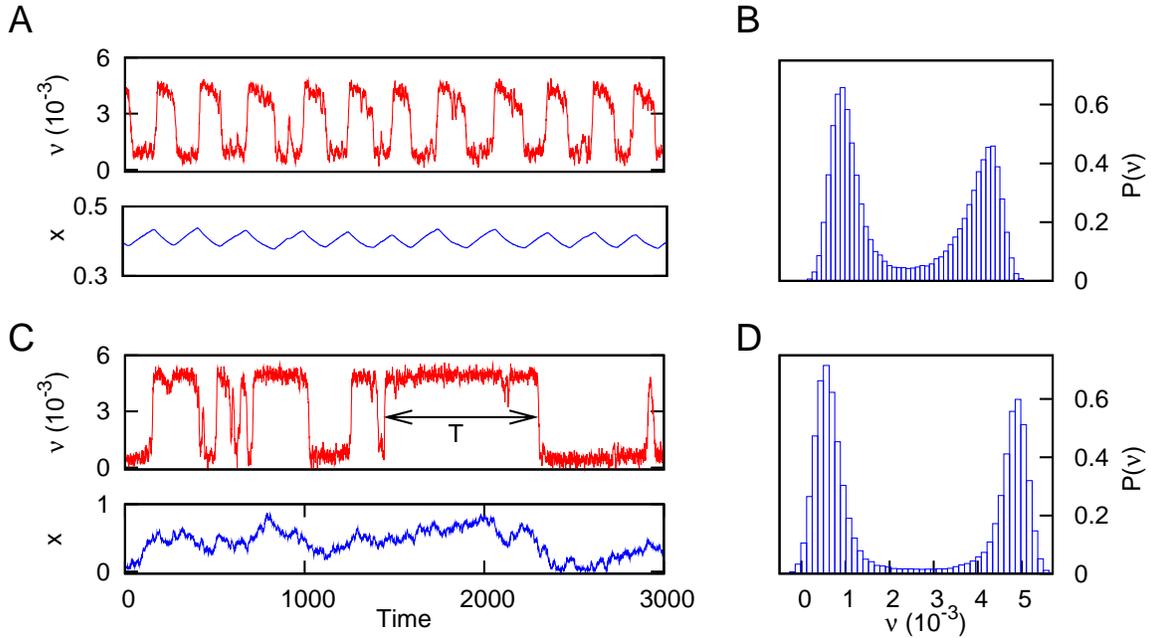}
\end{center}
\caption{
{\bf Time series showing the dynamics of our system.} (A) Time series of the mean firing rate of the neural population for deterministic depressing synapses. The temporal evolution of the variable $x$ is also plotted for illustration purposes. (B) Histogram of the mean firing rate, which shows the existence of two well defined states of activity in $\nu \sim 10^{-3}$ and $\nu \sim 5\cdot 10^{-3}$, corresponding to the down and up states respectively. The values of the parameters are $J=1.2~V,~\tau_r=1000,~u=0.6,~D=0,~\delta=0.3$ and $\nu_m=5\cdot10^{-3}$. (C) Same as (A), but with a certain level of intrinsic stochasticity on the dynamics of the synapses (concretely, we set $D=20$). The two-headed arrow shows a typical interval of permanence in the up state, denoted by $T$. (D) Same as (B), but for $D=20$. The other parameters take the same values as in (A) and (B).}
\label{fig2}
\end{figure}

\begin{figure}[!ht]
\begin{center}
\includegraphics[width=5in]{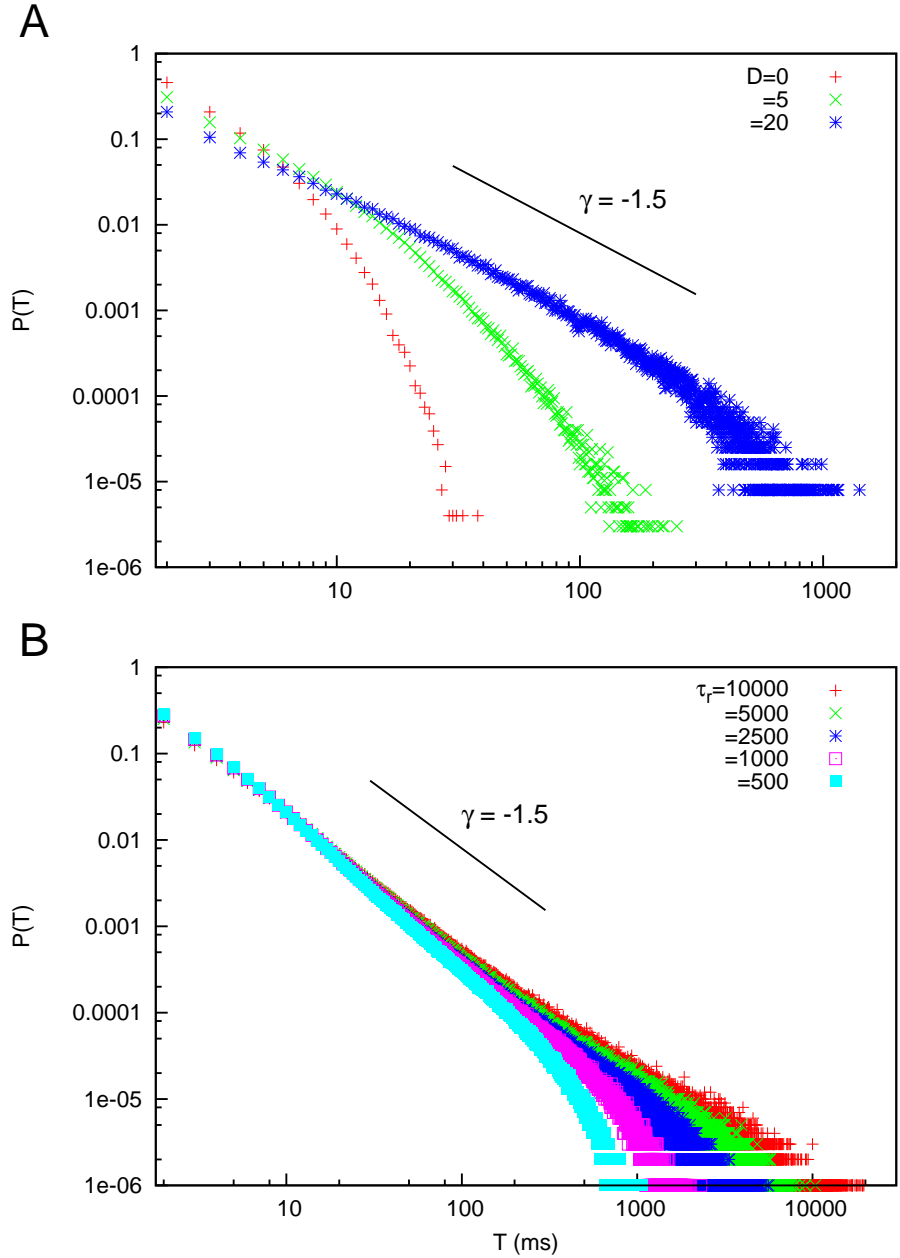}
\end{center}
\caption{
{\bf Probability distributions of permanence times in the up state.} (A) Probability distribution $P(T)$, obtained with numerical simulations, for different values of the noise strength $D$. One can see that high values of $D$ lead to the appearance of power-law distributions $P(T) \sim T^{-\gamma}$ with $\gamma =3/2$, as the mean-field solution predicts. For numerical simulations, we employed time series of duration $10^6$ and averaged over $100$ trials. The values of the other parameters were $J=1.1~V,~u=0.6,~\tau_r=1000,~\delta=0.3$ and $\nu_m=5\cdot 10^{-3}$. To compute $P(T)$, we have considered that the up state has been reached during a period $T$ (with $T>2$) if $\nu>\eta \nu_m$ during this period. We set $\eta=0.8$. (B) Probability distributions of permanence times in the up state, for different values of $\tau_r$ and fixed $\Delta x \simeq 0.065$ and $D/\tau_r =0.02$. In order to fix $\Delta x$ and $D/\tau_r$, we have conveniently modified $J$ and $D$, respectively, for each value of $\tau_r$. We employed time series of duration $10^6$ and averaged over $600$ trials. Other parameters are $u=0.04,~\delta=0.3$ and $\nu_m=5\cdot 10^{-3}$.}
\label{fig3}
\end{figure}

\begin{figure}[!ht]
\begin{center}
\includegraphics[width=6.5in]{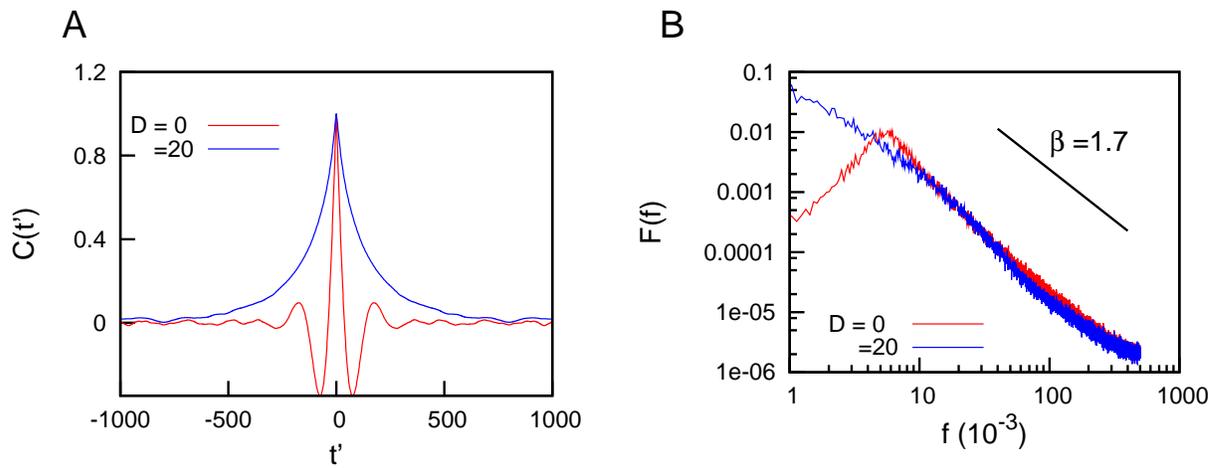}
\end{center}
\caption{
{\bf Autocorrelation and power spectra.} (A) Autocorrelation function of the mean firing rate for deterministic ($D=0$) and stochastic ($D=20$) synapses, in the presence of STD. (B) Power spectra of the mean firing rate for the two cases illustrated in (A). For both panels, we have averaged over $10^5$ time series of duration $10^6$ each, and we have fixed $J=1.1~V,~u=0.6,~\tau_r=1000,~\delta=0.3$ and $\nu_m=5\cdot 10^{-3}$.}
\label{fig4}
\end{figure}

\begin{figure}[!ht]
\begin{center}
\includegraphics[width=6in]{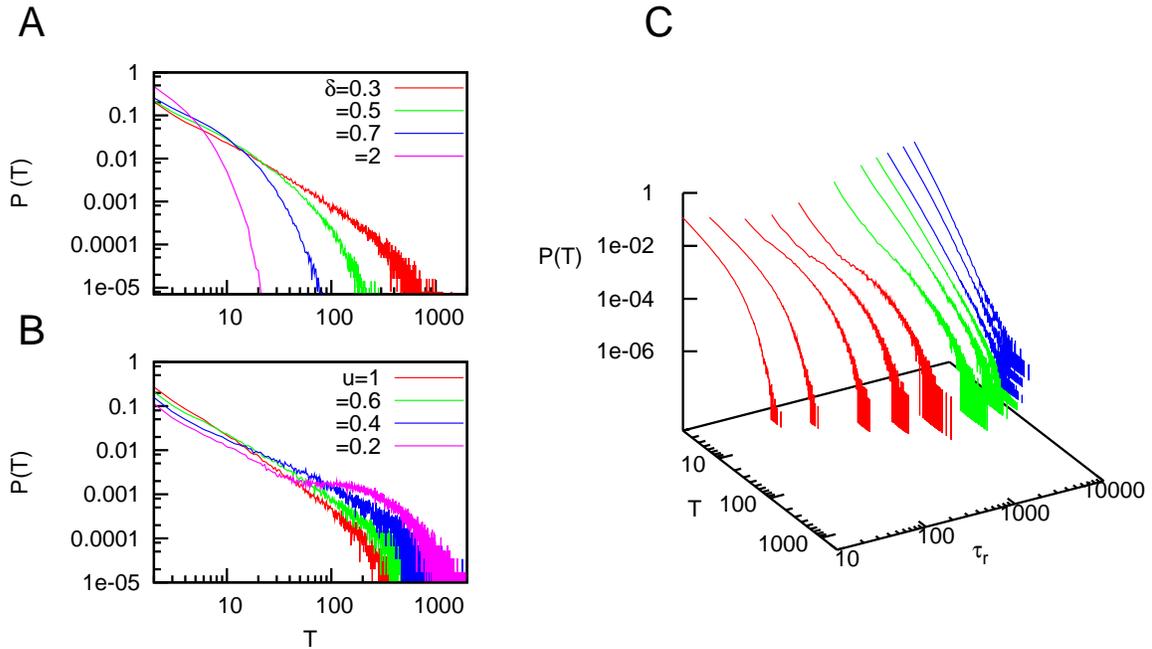}
\end{center}
\caption{
{\bf Influence of other parameters of the model.} (A) Probability distributions of permanence times in the up state, for different values of $\delta$. Other parameters are $J=1.1~V,~u=0.6,~\tau_r=1000,~D=20$ and $\nu_m=5\cdot 10^{-3}$. (B) Same as in (A), but for different values of $u$. The other parameters take the same values as in (A), except for $\delta=0.3$. (C) Probability distribution $P(T)$ as a function of $T$ and $\tau_r$. The three different regimes are shown with different colors (see main text for details). Other parameters are $J=1.1~V,~u=0.6,~D=20,~\delta=0.3$ and $\nu_m=5\cdot 10^{-3}$. For all panels, we have averaged over $100$ times series of duration $10^6$ each.}
\label{fig5}
\end{figure}

\begin{figure}[!ht]
\begin{center}
\includegraphics[width=6in]{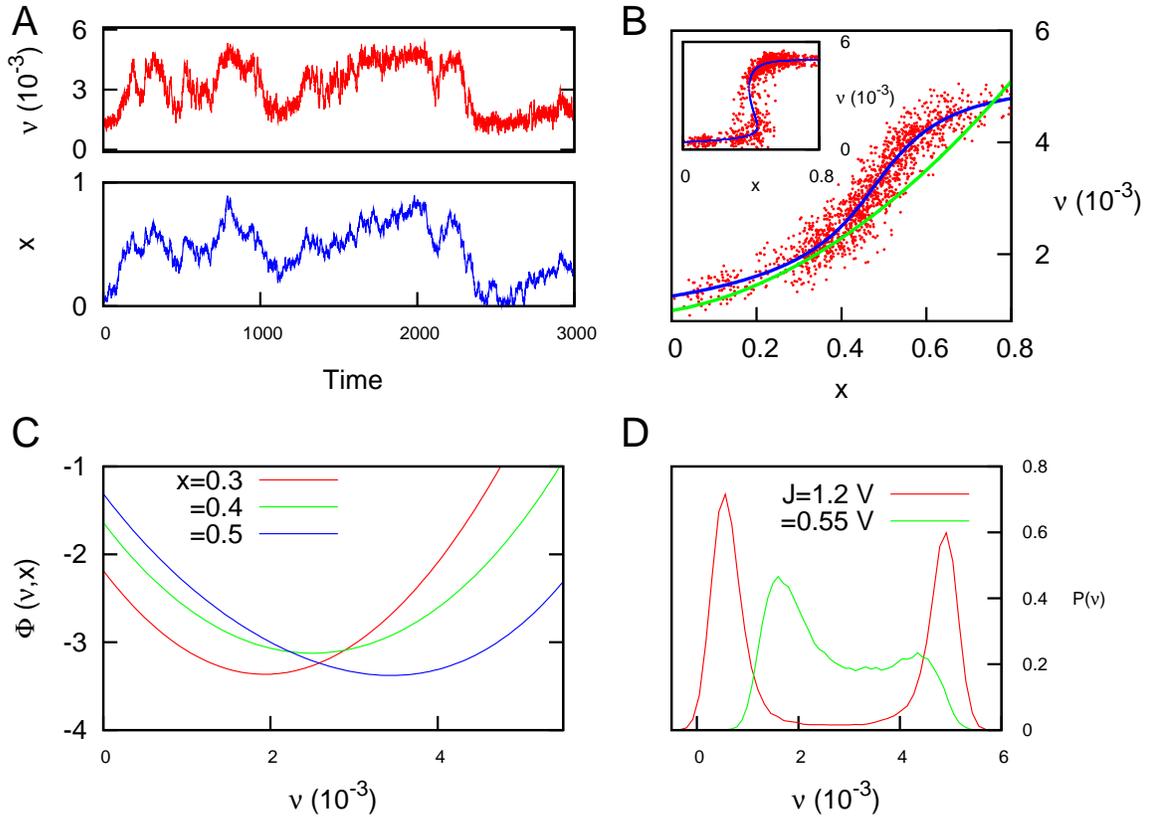}
\end{center}
\caption{
{\bf Behavior of the system when the condition $Jx_0\nu_0<1$ holds.} (A) Time series of the variables $\nu$ and $x.$ (B) The same time series, but represented on the $x-\nu$ plane, illustrates the fact that $\nu$ is a slave variable of $x$ (although some level of inner stochasticity on $\nu$ is still present). The green line corresponds to the approximate expression (\ref{approxnumono}), while the blue line is the numerical evaluation of the fixed point solutions of $\nu(t)$ (see equation (\ref{mfnu})). The inset shows the situation in which the system shows a bistable dynamics, analyzed in the previous section. (C) The potential function as a function of $\nu$ for different values of $x$. One can appreciate the existence of only one minimum, whose location is controlled by $x$. (D) Histograms of the mean firing rate of the system for different values of $J$. For the cases showed in this panel, the condition $J x_0 \nu_0<1$ is only satisfied for the case $J=0.55$. For all panels, $u=0.6,~\tau_r=1000,~D=20,~\delta=0.3,~\nu_m=5\cdot 10^{-3}$, and $J=0.55~V$ unless specifically specified.}
\label{fig6}
\end{figure}

\begin{figure}[!ht]
\begin{center}
\includegraphics[width=6.5in]{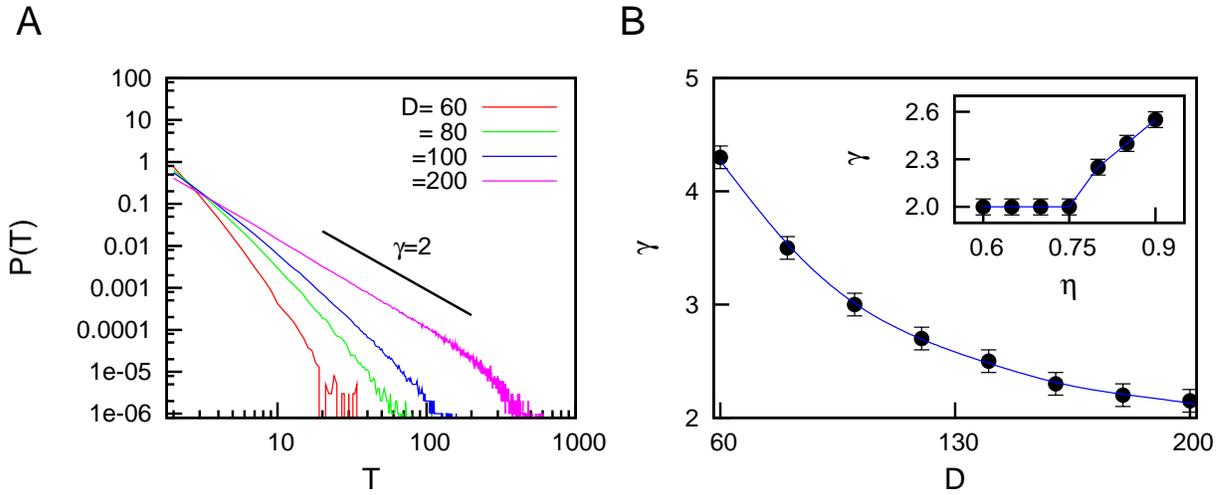}
\end{center}
\caption{
{\bf Statistics of permanence times in the up state for $Jx_0\nu_0<1$.} (A) Probability distribution of permanence times in the up state in the $J x_0 \nu_0<1$ regime, for $\eta=0.75$ and different values of $D$. One can see that power law relations $P(T) \sim T^{-\gamma}$ appear. (B) Dependence of $\gamma$ with $D$ for the conditions presented in (A). The inset shows the dependence of $\gamma$ with the parameter $\eta$ for the case $D=200$. We have averaged over $100$ time series of duration $10^6$ each. Other parameters are $J=0.05~V,~\tau_r=1000,~u=0.6,~\delta=0.3$ and $\nu_m=5\cdot 10^{-3}$.}
\label{fig7}
\end{figure}

\begin{figure}[!ht]
\begin{center}
\includegraphics[width=4in]{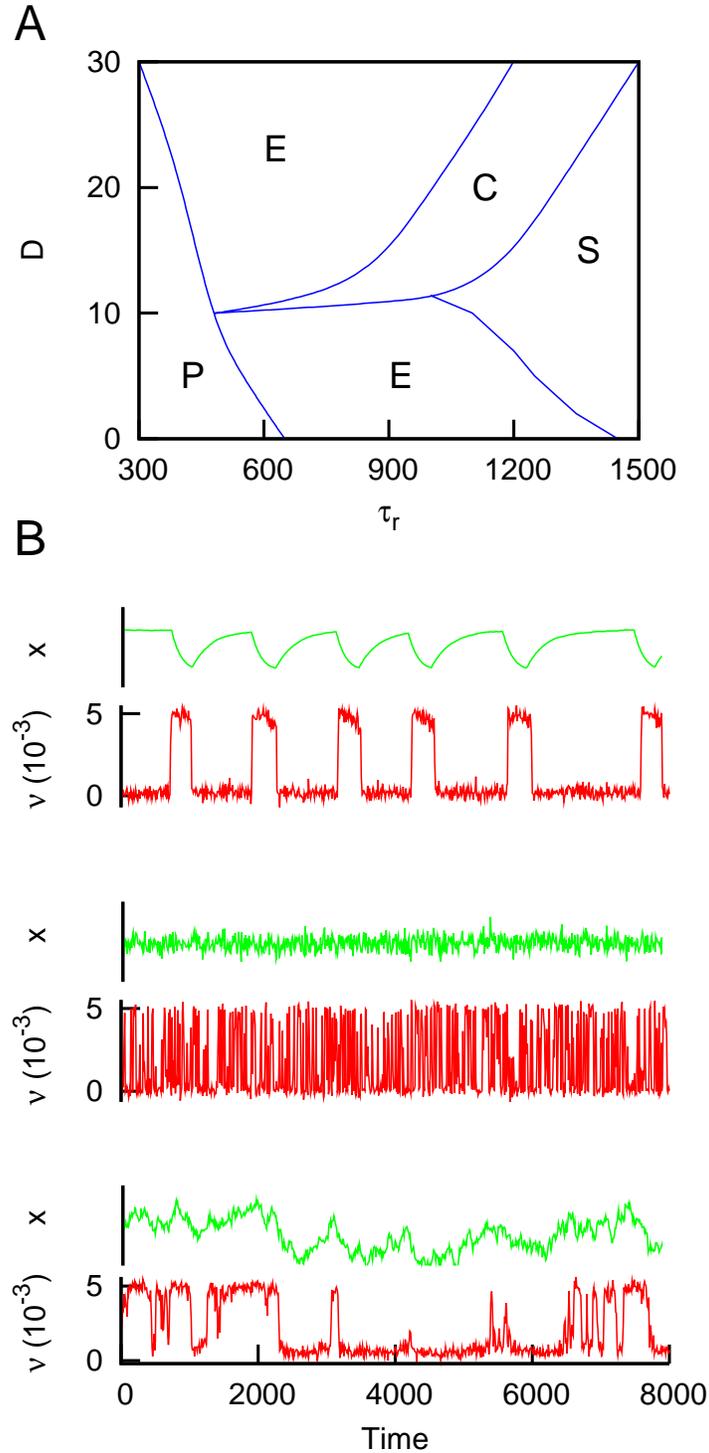}
\end{center}
\caption{
{\bf The different dynamical regimes of the model.} (A) Phase plot which shows the different behaviors found in our system. These behaviors corresponds to time series of $\nu$ for which permanence times in the up state follow an exponential distribution (E), a power-law distribution $P(T) \sim T^{-\gamma}$ with $\gamma=3/2$ (C), or a power-law distribution with $\gamma>3/2$ (S). In addition, a phase with a well-defined duration of the up state is found (P). In panel (B) some of these behaviors are depicted. From top to bottom one can see situations P, E and C. Other parameters are $J=1.1~V,~u=0.6,~\delta=0.3$ and $\nu_m=5\cdot 10^{-3}$.}
\label{fig8}
\end{figure}

\end{document}